\documentclass[prx,nopacs,twocolumn,aps]{revtex4-1}

\usepackage{amsmath}
\usepackage{bm}
\usepackage{epsfig}
\usepackage{amssymb}

\usepackage{graphicx}
\usepackage{epsfig}
\usepackage{psfrag}

\usepackage{cancel}
\usepackage{color}

\newcommand{\C}{\mathcal{C}}

\newcommand{\HH}{\mathbb{H}}
\newcommand{\PP}{\mathbb{P}}
\newcommand{\W}{\mathbb{W}}

\newcommand{\Tr}{\mathop{\mathrm{Tr}}}

\begin{document}

\title{Out-of-equilibrium evolution of kinetically constrained many-body quantum systems under purely dissipative dynamics}

\author{Beatriz Olmos}

\author{Igor Lesanovsky}

\author{Juan P. Garrahan}

\affiliation{School of Physics and Astronomy, University of Nottingham, Nottingham, NG7 2RD, UK}

\begin{abstract}
We explore the relaxation dynamics of quantum many-body systems that undergo purely dissipative dynamics through non-classical jump operators that can establish quantum coherence. Our goal is to shed light on the differences in the relaxation dynamics that arise in comparison to systems evolving via classical rate equations. In particular, we focus on a scenario where both quantum and classical dissipative evolution lead to a stationary state with the same values of diagonal or ``classical'' observables. As a basis for illustrating our ideas we use spin systems whose dynamics becomes correlated and complex due to dynamical constraints, inspired by kinetically constrained models (KCMs) of classical glasses. We show that in the quantum case the relaxation can be orders of magnitude slower than the classical one due to the presence of quantum coherences. Aspects of these idealized quantum KCMs become manifest in a strongly interacting Rydberg gas under electromagnetically induced transparency (EIT) conditions in an appropriate limit. Beyond revealing a link between this Rydberg gas and the rather abstract dissipative KCMs of quantum glassy systems, our study sheds light on the limitations of the use of classical rate equations for capturing the non-equilibrium behavior of this many-body system.
\end{abstract}


\maketitle

\section{Introduction}

The study of many-body quantum systems that undergo {\em purely dissipative} dynamics but at the same time feature quantum coherences and superpositions is an emerging theme that has attracted much interest in the past few years \cite{Kraus08,Diehl08,Diehl11,Bardyn12,Schindler13}. The catalyst was the realization that an appropriately engineered dissipative dynamics with non-classical jump operators represents a route towards the preparation of specific many-body states and non-equilibrium quantum phases with potential to be a resource for quantum computation \cite{Verstraete09}.

In this paper, we aim to study the role that quantum coherences and superpositions play in the relaxation of these dissipative many-body quantum systems. In particular, we will study systems with purely dissipative and Markovian dynamics generated by a Lindblad master equation formed by non-classical jump operators whose action can bring the system into quantum superpositions. The stationary state of these systems will be given by a pure state annihilated by all jump operators, an absorbing or ``dark'' state which in general will display quantum coherences. Moreover, these systems will be constructed such that in the stationary state the diagonal of the density matrix coincides with the equilibrium probability distribution of a completely classical rate equation. Our aim will be to contrast the relaxation given by this dissipative, yet quantum, dynamics with the one obtained via classical rate equations.

In order to tackle this task we will focus on a class of quantum spin systems inspired by kinetically constrained models of glasses (KCMs) \cite{Fredrickson1984,Jackle1991,[For reviews see: ]Ritort2003,*Garrahan2010}. These are classical stochastic models with trivial static equilibrium properties but complex collective dynamics due to the imposed kinetic constraints. Here we construct the purely dissipative quantum counterparts of these systems by ensuring that the values of diagonal observables (e.g. the density of excitations or the density-density correlations) in the stationary state coincide with the classical ones.  In classical KCMs all the complexity is found in the dynamics and not in their stationary state \cite{Ritort2003}.  The quantum dissipative KCMs we consider share this property, and are therefore a good test bed to study the differences and similarities in the relaxation between classical and purely dissipative quantum dynamics. A significant finding is that even when the relaxation timescales of the diagonal observables are similar, the coherences in the quantum system can take orders of magnitude longer to relax.
The models we construct can be regarded as an instance of quantum glasses, that is, interacting quantum systems whose real time relaxation is due to fluctuations of both thermal and quantum origin \cite{Markland11,Olmos12,Nussinov2012,Lechner13,Diaz-Mendez2014}.  Quantum glassiness is a timely topic also due to its relevance to thermalisation in quantum systems \cite{Rigol08,Polkovnikov2011} and to many-body localization \cite{Basko2006,Pal2010,Nandkishore2014}.

Finally, we show that aspects of these quantum KCMs, and therefore aspects of quantum glassiness more generally, become manifest in an ensemble of excited (Rydberg) atoms \cite{Gallagher84} under electromagnetically induced transparency (EIT) conditions. Our intention for drawing this connection is two-fold: Firstly, it demonstrates that the rather idealized features of quantum KCMs such as kinetic constraints and a purely dissipative, yet quantum, dynamics are actually present in a system currently studied by many experimental groups \cite{Pritchard10,Adams13,Schempp10,Schwarzkopf11,*Schwarzkopf13,Peyronel12,Maxwell13}. Secondly, this discussion sheds light on the limitations of the description of the dynamics of the EIT Rydberg gas in terms of classical rate equations, currently employed by numerous authors \cite{Ates06,Ates07-1,Ates07-2,Hoening13,Petrosyan13,Petrosyan13-1,Hoening14,Sanders14}.

\section{General setup}

Consider a classical stochastic system described by the master equation
\begin{equation}
  \partial_{t} |P(t)\rangle = {\mathbb W} |P(t) \rangle,
  \label{eq:MEC}
\end{equation}
where the vector $|P(t)\rangle \equiv \sum_{\C} P(\C;t) | \C \rangle$ represents the probability distribution at time $t$, and $\{ | \C \rangle \}$ is an orthonormal configuration basis. For a continuous time Markov chain the classical master operator reads,
\begin{equation}
{\mathbb W} \equiv \sum_{\C' \neq \C} W_{\C \to \C'} |\C' \rangle \langle \C|
- \sum_{\C} R_{\C} |\C \rangle \langle \C|,
\label{eq:W}
\end{equation}
where $W_{\C \to \C'}$ is the transition rate from $\C$ to $\C'$, and $R_{\C} \equiv \sum_{\C'} W_{\C \to \C'}$ the escape rate from the configuration $\C$. Let us assume that the dynamics obeys {\em detailed balance} with respect to the equilibrium probability $p_{\rm eq}(\C)$, i.e., the transition rates satisfy, $p_{\rm eq}(\C) W_{\C \to \C'} = p_{\rm eq}(\C') W_{\C' \to \C}$. This condition allows to transform the stochastic operator $\W$ into a Hermitian one $\HH$ through a similarity transformation,
\begin{equation*}
  \HH \equiv - \PP^{-1} \W \PP \quad\mathrm{with}\quad \PP \equiv \sum_{\C} \sqrt{p_{\rm eq}(\C)} | \C \rangle \langle \C |.
\end{equation*}
Note that the ground state $|{\rm g.s.}\rangle$ of $\HH$ (with eigenvalue 0) is directly related to the equilibrium probability as $|{\rm g.s.}\rangle \equiv \sum_{\C} \sqrt{p_{\rm eq}(\C)} | \C \rangle$.

Our aim is now to define a quantum model undergoing {\em purely dissipative quantum dynamics} generated by a quantum master equation of the Lindblad form \cite{Gardiner04},
\begin{equation}
  \partial_t\rho=\sum_\mu J_\mu \rho J_\mu^{\dagger}-\frac{1}{2}\left\{J_\mu^{\dagger} J_\mu,\rho\right\},
\label{eq:ME}
\end{equation}
such that the dynamics converges to a stationary state $\rho_{\rm s.s.}$ where the expectation value of any classical operator $\hat{O}$ (diagonal in the classical basis of configurations) is the same as in the classical equilibrium distribution
\begin{equation}
  \langle \hat{O} \rangle = \Tr[\hat{O} \rho_{\rm s.s.}] = \sum_{\C} p_{\rm eq}(\C) O(\C).
  \label{eq:O}
\end{equation}
This can be achieved by defining the following operators associated to the classical transitions between any pair of configurations $\mu = (\C,\C')$,
\begin{equation}
J_\mu \equiv | \psi \rangle \left( \sqrt{W_{\C \to \C'}} \langle \C | - \sqrt{W_{\C' \to \C}} \langle \C' | \right) ,
\label{eq:J}
\end{equation}
where $| \psi \rangle$ can in principle be any normalized state. One can show that the Hermitian form of the master operator can be written as
\begin{equation*}
  \HH =\frac{1}{2}\sum_{\mu} J_\mu^{\dagger} J_\mu.
\end{equation*}
Moreover, $J_\mu |{\rm g.s.}\rangle = 0$ for all $\mu$, that is, the state $|{\rm g.s.}\rangle$ is a {\em dark state} for all operators $J_\mu$. Thus, by considering the operators $J_\mu$ of Eq.\ (\ref{eq:J}) as quantum jump operators in Eq.\ (\ref{eq:ME}) the quantum dynamics converges to a pure stationary state $\rho_{\rm s.s.} = |{\rm g.s.}\rangle \langle {\rm g.s.}|$, where indeed the expectation values of classical operators correspond to the classical equilibrium ones as required in (\ref{eq:O}). The key question is what difference, if any, there is between the quantum dynamics defined by Eqs.\ (\ref{eq:ME}) and (\ref{eq:J}) and the classical one generated by (\ref{eq:MEC}) and (\ref{eq:W}).

\section{Quantum vs.\ classical dissipative dynamics in KCMs}

To gain a general understanding of this question we consider a particular set of quantum spin models based on classical KCMs \cite{Ritort2003}. The models we consider are defined in terms of $N$ binary variables on the sites of a lattice such that $\left|0_k\right>$ or $\left|1_k\right>$ corresponds to the $k$-th spin in the down or up state, respectively. The classical models can also be written in terms of a master equation like (\ref{eq:ME}) but with a pair of jump operators for each site $k=1\dots N$ that act on a diagonal density matrix and connect a classical configuration with another one that differs by the flip of the $k$-th spin,
\begin{equation}
  J_{k \uparrow} = \sqrt{\lambda\kappa} f_k \sigma_k^+\quad\mathrm{and}\quad J_{k \downarrow} =\sqrt{\lambda(1-\kappa)} f_k \sigma_k^-. \label{eq:classical_jump_ops}
\end{equation}
Here, $\lambda$ is the bare rate of jumps, $\kappa\in[0,1]$ is the average site occupation in equilibrium and $\sigma_k^\pm$ are the spin-$1/2$ ladder operators. The operator $f_{k}$ represents a {\em kinetic constraint} on site $k$, that is, a function of diagonal operators $n_j=\sigma_j^+\sigma_j^-$ $\forall \; j\neq k$, which conditions the dynamics of site $k$ depending on the state of its neighbours.

In contrast, in the quantum model there is a \emph{single jump operator per site}, which establishes superpositions between $\left|0_{k}\right>$ and $\left|1_{k}\right>$:
\begin{equation}
J_k \equiv \sqrt{\lambda} ~ U_k ~ \left|B_{k}\right>\left<B_{k}\right| ~ f_k ,
\label{eq:JJ}
\end{equation}
where
\begin{equation*}
  \left|B_{k}\right> = \sqrt{\kappa}\left|0_{k}\right> - \sqrt{1-\kappa}\left|1_{k}\right>.
\end{equation*}
It is important to note that the construction of the quantum problem allows more freedom than the classical case: The relaxation is not only controlled by the constraint function $f_k$, but also by an additional arbitrary set of unitaries $U_k$ (discussed in detail later). However, similarly to the classical case, the stationary state of the dynamics generated by (\ref{eq:ME}) is, independently of the choices for $U_k$ and $f_k$, the same pure state of the direct product form
\begin{equation*}
  \rho_\mathrm{s.s.}=\bigotimes_{k}|S_{k}\rangle\langle S_{k}|,
\end{equation*}
where
\begin{equation*}
  |S_{k}\rangle = \sqrt{1-\kappa} \left|0_{k}\right> + \sqrt{\kappa} \left|1_{k}\right>,
\end{equation*}
so that $\langle S_{k}| B_{k} \rangle=0$. Furthermore, the quantum problem is constructed such that the expectation values of all diagonal operators in the $|0,1\rangle$ basis (such as the density of excitations or the density-density correlations) in the stationary state $\rho_\mathrm{s.s.}$ \emph{coincide} with those at equilibrium in the classical one.

As a first elementary comparison we can consider an unconstrained single spin ($f=1$). Its classical dynamics is determined by Eq.\ (\ref{eq:ME}) with the two jump operators $J_\uparrow=\sqrt{\lambda \kappa}\sigma^+$ and $J_\downarrow=\sqrt{\lambda(1-\kappa)}\sigma^-$. In this case, the relaxation of the density of excitations $\langle n(t) \rangle$ is exponential with a single timescale given by $\tau_{\rm cl} = \lambda^{-1}$. The quantum model has a single jump operator of the form (\ref{eq:JJ}), $J=\sqrt{\lambda} ~U \left|B\right>\left<B\right|$. For the purpose of illustration we choose the free unitary $U$ to be a spin rotation around the $y$-axis, i.e., $U=\exp{(i\theta\sigma^y)}$ with the angle $0\leq\theta\leq\pi$. The solution of the master equation reveals now two relaxation timescales for the density of excitations, $\tau_{\rm q} = 2 \lambda^{-1}$ and $\tau'_{\rm q} =  (\sin{\theta})^{-2} \lambda^{-1}$. The emergence of the second $\theta$-dependent timescale, which is due to $U$ and hence absent in the classical case, can be understood by unravelling the dynamics in terms of quantum jump trajectories with respect to the jump operators $J$ \cite{Molmer93,Dalibard92}. Here, each trajectory is obtained by a ``no-jump'' non-Hermitian time-evolution of the system via $H_\mathrm{eff}=-iJ^\dag J/2$ interspersed with quantum jumps. While the evolution between jumps is $U$-independent and thus $\theta$-independent, each quantum jump may take the system closer or further away from the stationary state $|S\rangle$ depending on the angle $\theta$. In particular, when $\theta=\pi/2$ the time $\tau'_{\rm q}$ coincides with $\tau_{\rm cl}$. The limit of $\theta=0$ is the opposite extreme. Here, the jump operators are Hermitian and the completely mixed state (proportional to the identity) becomes a further stationary state together with the pure one. The system then relaxes into a combination of the pure state and the completely mixed state that depends on the initial conditions.

\begin{figure}[t!]
  \includegraphics[width=\columnwidth]{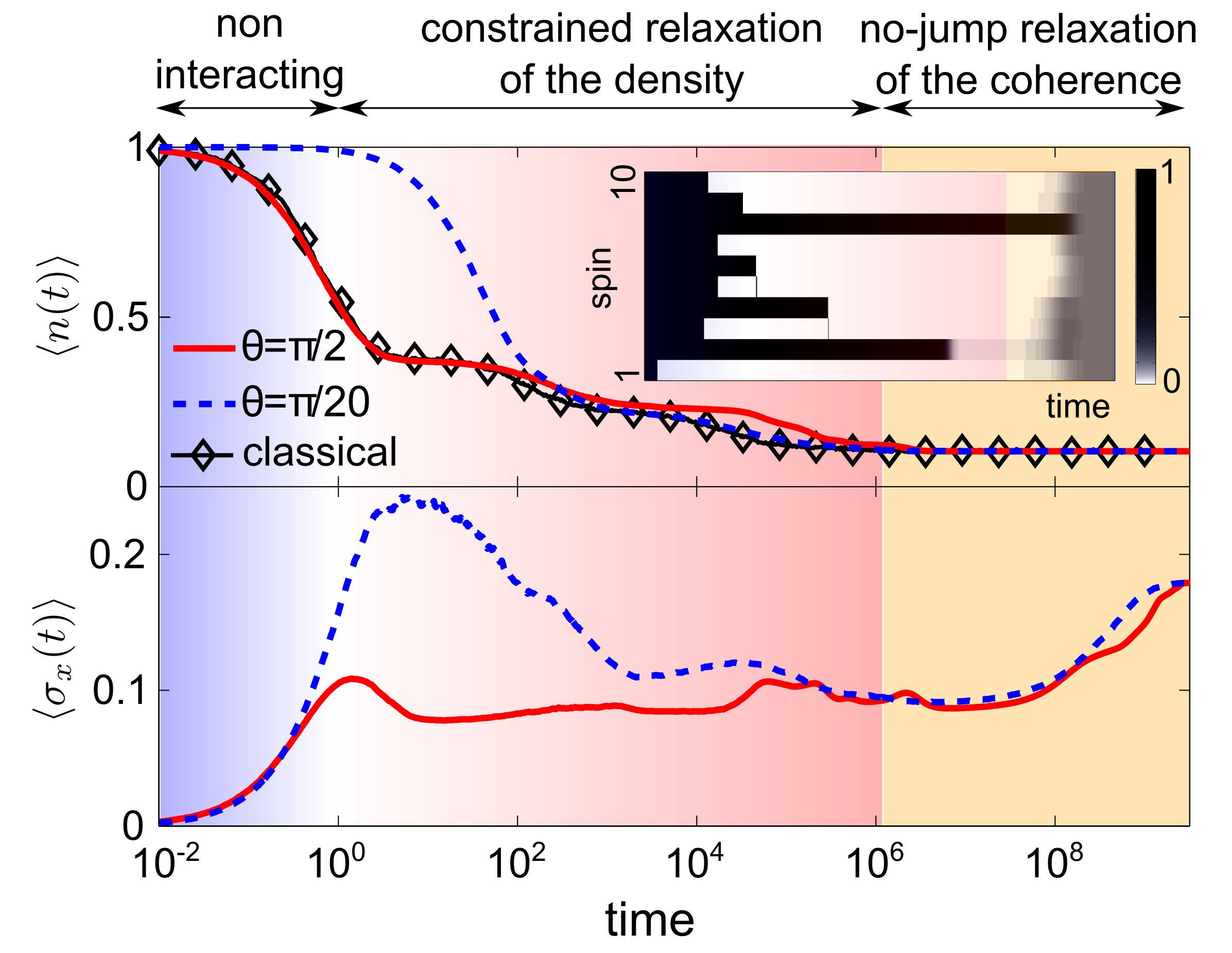}
  \caption{Relaxation of both classical and quantum dissipative East models for $\kappa/(1-\kappa)=10^{-2}$ and a system size of $N=10$ spins (time in units of $\lambda^{-1}$). The upper panel shows the average excitation density, $\langle n(t) \rangle$. The lower panel shows the evolution of the coherences, represented by $\langle \sigma^x (t) \rangle$, in the quantum case. Inset: time-evolution of $\langle n_k \rangle$ for $k=1\dots N$ in a single trajectory for $\theta=\pi/2$.}
  \label{fig:East}
\end{figure}

We now turn to the actual constrained many-body models \footnote{Note, that there are other ways to define quantum versions of these KCMs which are not purely dissipative \cite{Olmos12}.}. We will consider here two models determined by the kinetic constraints: $f_{k}^{\rm East} = n_{k+1}$, which allows transitions at site $k$ only if site $k+1$ has a projection on the spin state $\left|1\right>$ (in analogy with the classical East model \cite{Jackle1991,Ritort2003,Chleboun2013}), and $f_{k}^{\rm FA} = n_{k+1} + n_{k-1} - n_{k+1} n_{k-1}$, which allows transitions at site $k$ if and only if at least one of the sites at $k\pm1$ has a projection on $\left|1\right>$ (in analogy with the classical Fredrickson-Andersen (FA) model \cite{Fredrickson1984,Ritort2003}).

We first focus on the East model. When $\kappa$ is small the system's relaxation encounters a conflict between the probability cost of flipping spins up, and the need for excited sites to facilitate neighbouring ones through the constraint $f_{k}^{\rm East}$. In the classical model this gives rise to hierarchical dynamics \cite{Ritort2003,Chleboun2013}, which manifests for example in metastable plateaus in the relaxation of the density, see Fig. \ref{fig:East}. For the quantum counterpart, Eqs.\ (\ref{eq:ME},\ref{eq:JJ}), the dynamics depends on the angle $\theta$ that defines the unitaries $U_k=\exp{(i\theta\sigma^y_k)}$, which for simplicity we consider again to be local spin rotations. We use quantum jump Monte Carlo simulations to study the time evolution of the system numerically starting from a state of maximum density, see Fig. \ref{fig:East}.

\begin{figure}[t!]
  \includegraphics[width=0.9\columnwidth]{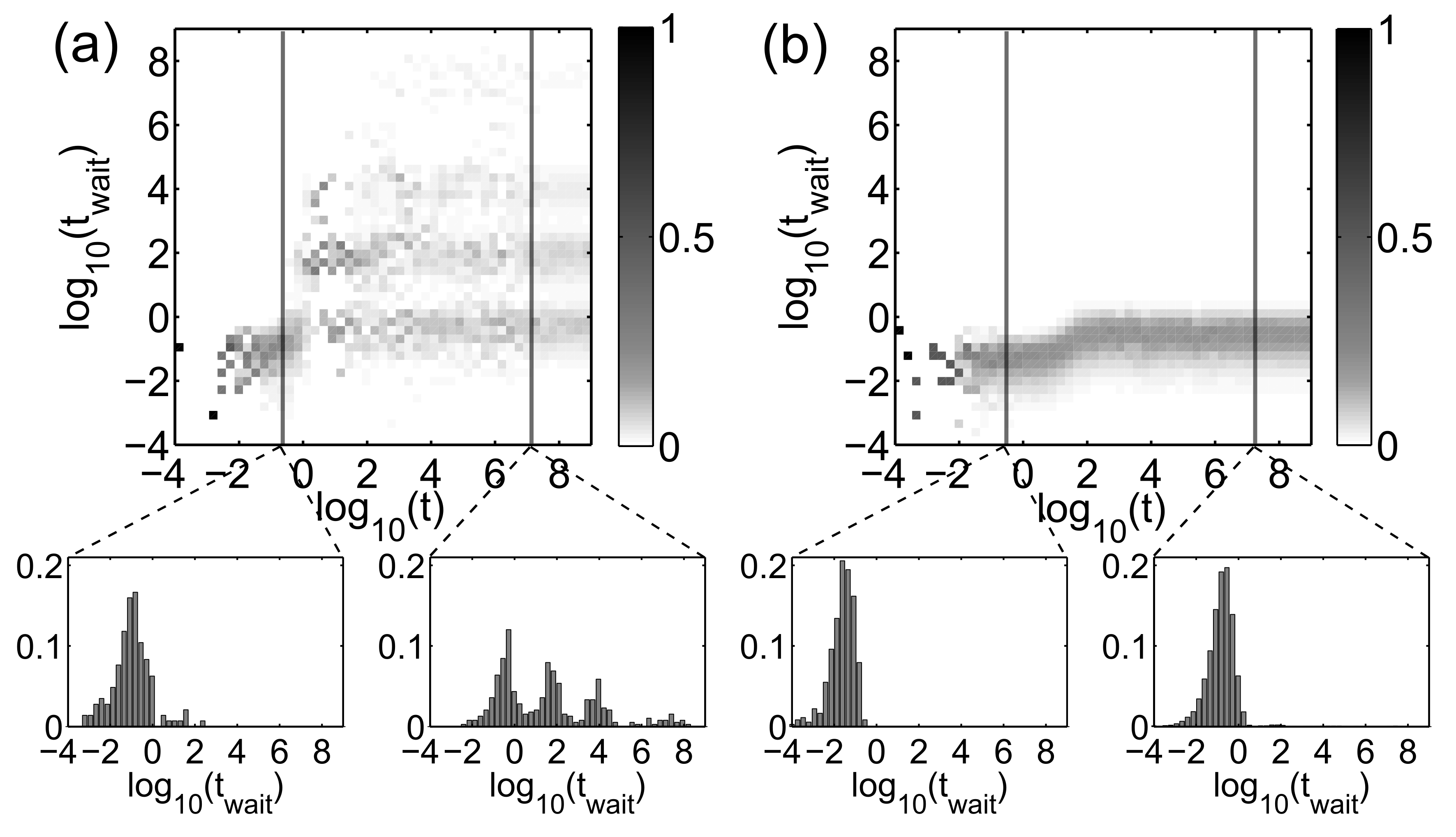}
  \caption{Waiting time distributions $t_\mathrm{wait}$ as a function of time in the dissipative quantum East model for $N=10$ and $\kappa/(1-\kappa)=10^{-2}$ (all times in units of $\lambda^{-1}$). (a) When $\theta=\pi/2$, the plateaus in the excitation density give rise to the appearance of several peaks in the waiting time distributions. (b) When $\theta=\pi/20$, every jump brings the system away from equilibrium making the occurrence of a subsequent jump more likely.}
  \label{fig:Times}
\end{figure}

One can distinguish several regimes in the quantum relaxation of the average density of excited sites. For short times one has effectively unconstrained dynamics, as the density of excitations is high and thus the constraints still do not play a role. When $\theta=\pi/2$, both the ``no jump" dynamics due to $H_\mathrm{eff}$ and the action of the non-classical jump operators help the system to reach a configuration where the excitations are isolated (see inset to Fig.\ \ref{fig:East}). Further relaxation needs excitations to effectively propagate, to meet and coalesce with others. This slow propagation leads to the different plateaus, in analogy with the classical East model \cite{Ritort2003,Sollich99}. In contrast, when $\theta \gtrsim 0$ every jump brings the system away from the stationary state, making another jump likely to occur. This different behavior between $\theta = \pi/2$ and $\theta \gtrsim 0$ becomes clear in the distributions of waiting times between jumps, see Fig.\ \ref{fig:Times}.

The key aspect in the dynamics of the quantum version of the East model is in the relaxation of the coherences, represented in the lower panel of Fig.\ \ref{fig:East} by $\langle \sigma^x(t) \rangle=1/N\sum_k\langle\sigma^x_k\rangle$. For any angle $\theta$ this observable takes approximately three orders of magnitude longer to relax than the density of excitations. The reason for this mismatch between the two timescales can be understood by looking into single trajectories (e.g. inset of Fig.\ \ref{fig:East}). Here, we observe that the strongly constrained propagation-coalescence relaxation of the density leaves the system eventually in a very inhomogeneous configuration with an isolated excitation, where the density profile is roughly $\dots \kappa \kappa 1 \kappa \kappa \dots$. While the overall value of the density of excitations coincides at this point with the stationary one, the last isolated excitation must be distributed or delocalized through the lattice in order to yield the "correct" stationary value for the coherences. Due to the highly constrained nature of the dynamics this takes orders of magnitude longer.

\begin{figure}[ht]
  \includegraphics[width=\columnwidth]{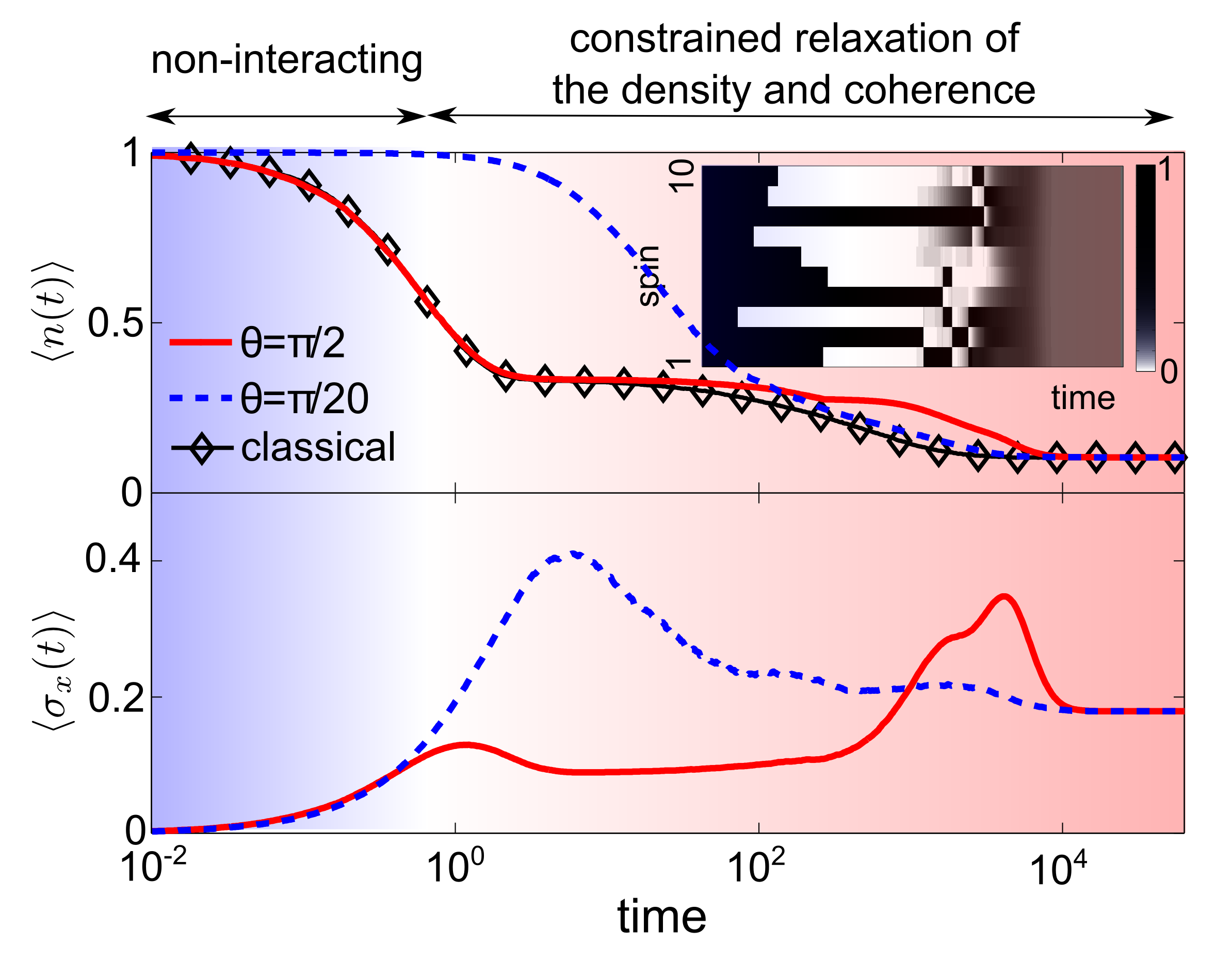}
  \caption{Relaxation of both classical and quantum dissipative FA models for $\kappa/(1-\kappa)=10^{-2}$ and a system size of $N=10$ spins (time in units of $\lambda^{-1}$). The upper panel shows the average excitation density, $\langle n(t) \rangle$. The lower panel shows the evolution of the coherences, represented by $\langle \sigma^x (t) \rangle$, in the quantum case. Inset: time-evolution of $\langle n_k \rangle$ for $k=1\dots N$ in a single trajectory for $\theta=\pi/2$.}\label{fig:FA}
\end{figure}

In other quantum models with less restrictive constraints, such as the quantum counterpart of the FA model, diagonal and off-diagonal observables are able to relax simultaneously (see Fig. \ref{fig:FA}). The reason is that due to the bidirectional nature of the constraints $f_{k}^{\rm FA}$ isolated excitations become delocalized during the relaxation (see inset of Fig.\ \ref{fig:FA}). In contrast with the East model, this leads quickly to an uniform density profile (i.e. each spin is in the same state, as established by $\rho_\mathrm{s.s.}$) which in turn means that also the coherences have reached their stationary value.

\section{Connection to a strongly interacting Rydberg gas}

In order to lift the abstract and seemingly artificial character of the current discussion we will now establish a link to a quantum-optical system, showing that features of the quantum KCMs indeed emerge here rather naturally. Specifically, we consider a gas of interacting Rydberg atoms under EIT conditions, which is currently in the focus of numerous theoretical \cite{Ates06,*Ates07-1,*Ates07-2,Hoening13,Petrosyan13,Petrosyan13-1,Hoening14,Sanders14} and experimental \cite{Pritchard10,Adams13,Schempp10,Schwarzkopf11,*Schwarzkopf13,Peyronel12,Maxwell13} investigations. We show below that the dynamics of this many-body system can indeed be described by a quantum master equation (\ref{eq:ME}) with non-classical jump operators which, in some limit, reduce to those of a quantum KCM very similar to Eq. (\ref{eq:JJ}).

\begin{figure}[t!]
\includegraphics[width=\columnwidth]{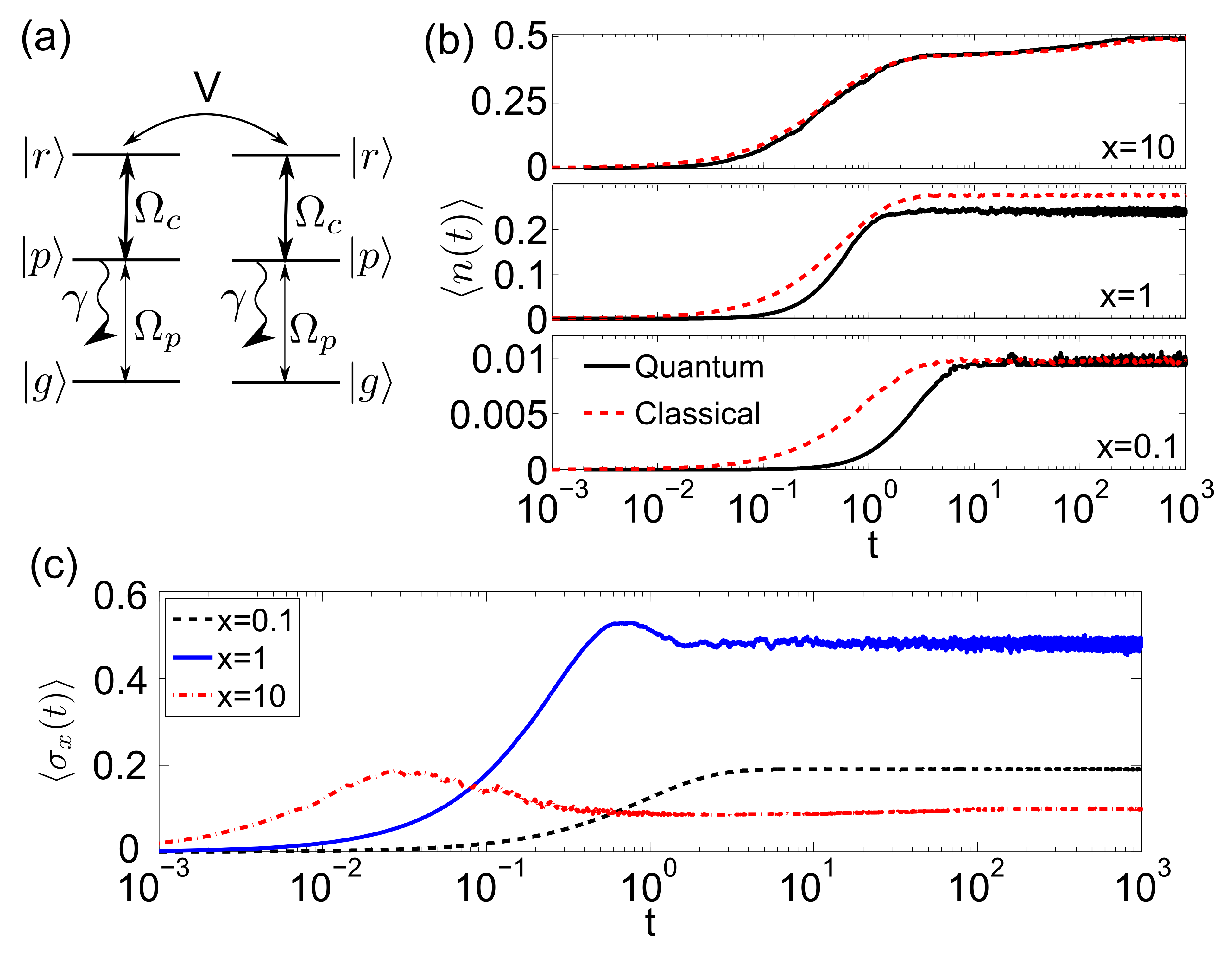}
  \caption{(a) Interacting atoms in a EIT configuration. Neighbouring atoms in the Rydberg state, $\left|r\right>$, interact with strength $V$. Strong decay, $\gamma$, of the intermediate level, $\left|p\right>$, effectively gives rise to purely dissipative quantum dynamics. (b) Relaxation of the expectation value of the excitation density of the classical excluded volume model (red, dashed lines) and the quantum Rydberg lattice gas (black, solid lines) for $x=\Omega_\mathrm{p}/\Omega_\mathrm{c}=0.1,1$ and $10$ (system size $N=8$). (c) Relaxation of the expectation value of the coherences, represented by $\langle \sigma^x (t) \rangle$, for the same three values of the parameter $x$.}
\label{fig:Realclas}
\end{figure}

The specific setting we have in mind consists of atoms trapped in a one-dimensional lattice, which are laser driven under EIT conditions as shown in Fig. \ref{fig:Realclas}(a), i.e. resonantly excited from the ground $\left|g\right>$ to the Rydberg state $\left|r\right>$ via an intermediate state $\left|p\right>$ with decay rate $\gamma$. When two neighboring atoms are simultaneously in $\left|r\right>$, they interact with interaction strength $V$. These dynamics are governed by the quantum master equation
\begin{equation*}
  \partial_t\rho={\cal L}_0\rho+{\cal L}_1\rho+{\cal L}_2\rho
\end{equation*}
with
\begin{eqnarray*}
  {\cal L}_0\rho&=&-iV\sum_k\left[\left|r_k\right>\!\left<r_k\right| \otimes\left|r_{k+1}\right>\!\left<r_{k+1}\right|,\rho\right]\\
  {\cal L}_1\rho&=&\gamma\sum_k\left(\left|g_k\right>\!\left<p_k\right|\rho \left|p_k\right>\!\left<g_k\right|- \frac{1}{2}\left\{\left|p_k\right>\!\left<p_k\right|,\rho\right\}\right)\\
  {\cal L}_2\rho&=&-i\sum_k\left[-\Omega_\mathrm{c}\left|p_k\right>\!\left<r_k\right|
+\Omega_\mathrm{p}\left|g_k\right>\!\left<p_k\right|+\mathrm{h.c.},\rho\right]
\end{eqnarray*}
with $\Omega_\mathrm{c}$ and $\Omega_\mathrm{p}$ being the Rabi frequencies of the lasers that couple $\left|p\right>$ to $\left|r\right>$ and $\left|g\right>$ to $\left|p\right>$, respectively.

An effective master equation of the form (\ref{eq:ME}) can be derived in the limits of $V\to\infty$ \cite{Lesanovsky11,Petrosyan13} and large but finite decay rate $\gamma$. By adiabatically eliminating the intermediate state $\left|p\right>$ one obtains an effective perturbative dynamics (following the procedure of Refs.\ \cite{Lesanovsky13,Marcuzzi14}). This reduces the problem to an ensemble of two-level systems that undergo a purely dissipative master equation (\ref{eq:ME}) with scaled time $t\rightarrow t(4\Omega_\mathrm{c}^2)/\gamma$ and non-classical jump operators
\begin{equation}
J^\mathrm{Ryd}_k=xp_k-p_{k-1}\sigma_k^-p_{k+1}, \label{eq:Ryd_quant}
\end{equation}
where we have defined $x=\Omega_p/\Omega_c$, $p_k=\left|g_k\right>\!\left<g_k\right|$ and $\sigma_k^-=\left|g_k\right>\!\left<r_k\right|$. In the limit $x\ll1$, the jump operators yield $J_k^\mathrm{Ryd} \approx \sqrt{\lambda}\left|0_k\right>\!\left<B_k\right|p_{k-1}p_{k+1}$, that is, in this limit the interacting many-body system realises an instance of the previously discussed quantum KCMs with constraint function $f_k=p_{k-1}p_{k+1}$, where $\kappa/(1-\kappa)=x^2$.

Beyond this conceptually interesting link, these results allow us to comment on the appropriateness of current theoretical efforts \cite{Ates06,*Ates07-1,*Ates07-2,Hoening13,Petrosyan13,Petrosyan13-1,Hoening14,Sanders14} that aim at capturing the dynamics of EIT Rydberg gases with classical rate equations. Some of them employ a description of the dynamics in terms of a classical KCM with constraint function
\begin{eqnarray*}
  f^\mathrm{Ryd}_k=p_{k-1}p_{k+1},
\end{eqnarray*}
which corresponds to modelling the excited Rydberg atoms by hard rods with an \emph{excluded volume} \cite{Petrosyan13}. A general problem appears to be that in these model descriptions only the ratio of the (classical) excitation and de-excitation rates, which enter Eqs. (\ref{eq:classical_jump_ops}) through the parameter $\kappa$, is given. This is not problematic when when one is interested only in stationary state properties, as e.g. in Ref. \cite{Petrosyan13}. However, in order to make a meaningful comparison of the dynamics between the classical excluded volume model and the effective dissipative quantum evolution defined by the jump operators (\ref{eq:Ryd_quant}) --- which is derived from first principles --- we need to find the appropriate absolute timescales, which depend on the choice of the rate $\lambda$. We do this "matching of time scales" by considering a single spin. In the classical case we find that its excitation density evolves as
\begin{equation*}
  \langle n_c(t)\rangle=\frac{x^2}{1+x^2}\left[1-e^{-t\lambda}\right],
\end{equation*}
which has been calculated by using the jump operators (\ref{eq:classical_jump_ops}) with $\kappa=x^2/(1+x^2)$. However, in the quantum case we obtain
\begin{eqnarray*}
  \langle n_q(t)\rangle=\frac{x^2}{1+x^2}\left[1+\frac{1+x^2}{1-x^2}e^{-t}- \frac{2}{1-x^2}e^{-t\frac{1+x^2}{2}}\right],
\end{eqnarray*}
which in the limit of $x\gg1$ yields
\begin{equation*}
  \langle n_q(t)\rangle \approx \frac{x^2}{1+x^2}\left[1-e^{-t}\right].
\end{equation*}
Setting $\lambda=1$ makes the two expressions match for short times and hence one might think of this as being an adequate absolute timescale. This choice is used in all simulations shown in Fig. \ref{fig:Realclas}(b) and (c).

As shown in Fig.\ \ref{fig:Realclas}(b), for certain parameters the stationary value of classical observables (here the excitation density) is indeed well reproduced by the classical hard objects model, but not necessarily \emph{the dynamics towards it}. For example, in the limit $x\ll1$ ($x=0.1$ in the figure), $J^{\mathrm{Ryd}}_k$ has the form (\ref{eq:JJ}) and thus, as shown earlier, the diagonal observables in the stationary state of the quantum system coincide with the ones given by the classical excluded volume model. However, the two dynamics are qualitatively different. For $x=10$ both the dynamics and the stationary expectation value of the excitation density coincide. This is in agreement with the results in Refs. \cite{Ates06,*Ates07-1,*Ates07-2}, where it was shown that indeed in the limit $x\gg1$ the classical excluded volume model yields a good approximation to the many-body quantum dynamics. Away from the two limits $x \ll 1$ and $x \gg 1$ the classical model reproduces neither the stationary expectation values of the diagonal observables nor the relaxation dynamics towards them.
Finally, let us investigate the time evolution of quantum observables. As an example we choose to monitor the expectation value of the single atom coherence, $\sigma_x$. In Fig. \ref{fig:Realclas}(c) one can see that $\langle\sigma_x\rangle$ relaxes simultaneously with the diagonal observables as in the FA model. It is important to note, however, that except in the limit $x\gg1$ the stationary state values of the coherence are not negligible, even being close to the possible maximal value $1/2$ for $x=1$. This is a further feature that is intrinsically impossible to capture by classical rate equations.

\section{Summary and Conclusions}
The aim of this paper was to shed light on the out-of-equilibrium relaxation dynamics of many-body systems that relax under a purely dissipative quantum dynamics. After some general considerations we have introduced and analyzed quantum versions of classical KCMs. Here we have found that coherences under certain circumstances can exhibit relaxation timescales that are orders of magnitude longer than those of classical observables. The discussed quantum generalizations of the KCMs are furthermore interesting because they can be thought of model systems for quantum glasses. Finally, we have established a link from the discussed seemingly abstract KCMs to the currently much studied system of Rydberg gases under EIT conditions. Here we have shown that they implement (in some limit) rather naturally an instance of a quantum KCM. This insight allowed us furthermore to comment on the appropriateness of classical rate equation models that are currently widely employed for studying the dynamics and statics of this many-body system.

The discussion presented in this work touches very general questions concerning the role of quantum effects in the relaxation of (glassy) many-body systems \cite{Olmos12,Markland11,Lechner13}. It also links to current experiments in the field of quantum optics and recent efforts in the domain of dissipative quantum state preparation, e.g. strongly correlated states of fermions \cite{Diehl10,Yi12}.

\acknowledgments
The work was supported in part by EPSRC Grant no. EP/I017828/1. The research leading to these results has received funding from the European Research Council under the European Union's Seventh Framework Programme (FP/2007-2013) / ERC Grant Agreement No. 335266 (ESCQUMA) and the EU-FET Grant No. 512862 (HAIRS). B.O. acknowledges funding by the University of Nottingham. We gratefully acknowledge discussions with Matteo Marcuzzi.


\begin{thebibliography}{47}%
\makeatletter
\providecommand \@ifxundefined [1]{%
 \@ifx{#1\undefined}
}%
\providecommand \@ifnum [1]{%
 \ifnum #1\expandafter \@firstoftwo
 \else \expandafter \@secondoftwo
 \fi
}%
\providecommand \@ifx [1]{%
 \ifx #1\expandafter \@firstoftwo
 \else \expandafter \@secondoftwo
 \fi
}%
\providecommand \natexlab [1]{#1}%
\providecommand \enquote  [1]{``#1''}%
\providecommand \bibnamefont  [1]{#1}%
\providecommand \bibfnamefont [1]{#1}%
\providecommand \citenamefont [1]{#1}%
\providecommand \href@noop [0]{\@secondoftwo}%
\providecommand \href [0]{\begingroup \@sanitize@url \@href}%
\providecommand \@href[1]{\@@startlink{#1}\@@href}%
\providecommand \@@href[1]{\endgroup#1\@@endlink}%
\providecommand \@sanitize@url [0]{\catcode `\\12\catcode `\$12\catcode
  `\&12\catcode `\#12\catcode `\^12\catcode `\_12\catcode `\%12\relax}%
\providecommand \@@startlink[1]{}%
\providecommand \@@endlink[0]{}%
\providecommand \url  [0]{\begingroup\@sanitize@url \@url }%
\providecommand \@url [1]{\endgroup\@href {#1}{\urlprefix }}%
\providecommand \urlprefix  [0]{URL }%
\providecommand \Eprint [0]{\href }%
\providecommand \doibase [0]{http://dx.doi.org/}%
\providecommand \selectlanguage [0]{\@gobble}%
\providecommand \bibinfo  [0]{\@secondoftwo}%
\providecommand \bibfield  [0]{\@secondoftwo}%
\providecommand \translation [1]{[#1]}%
\providecommand \BibitemOpen [0]{}%
\providecommand \bibitemStop [0]{}%
\providecommand \bibitemNoStop [0]{.\EOS\space}%
\providecommand \EOS [0]{\spacefactor3000\relax}%
\providecommand \BibitemShut  [1]{\csname bibitem#1\endcsname}%
\let\auto@bib@innerbib\@empty
\bibitem [{\citenamefont {Kraus}\ \emph {et~al.}(2008)\citenamefont {Kraus},
  \citenamefont {B\"uchler}, \citenamefont {Diehl}, \citenamefont {Kantian},
  \citenamefont {Micheli},\ and\ \citenamefont {Zoller}}]{Kraus08}%
  \BibitemOpen
  \bibfield  {author} {\bibinfo {author} {\bibfnamefont {B.}~\bibnamefont
  {Kraus}}, \bibinfo {author} {\bibfnamefont {H.~P.}\ \bibnamefont
  {B\"uchler}}, \bibinfo {author} {\bibfnamefont {S.}~\bibnamefont {Diehl}},
  \bibinfo {author} {\bibfnamefont {A.}~\bibnamefont {Kantian}}, \bibinfo
  {author} {\bibfnamefont {A.}~\bibnamefont {Micheli}}, \ and\ \bibinfo
  {author} {\bibfnamefont {P.}~\bibnamefont {Zoller}},\ }\href@noop {}
  {\bibfield  {journal} {\bibinfo  {journal} {Phys. Rev. A}\ }\textbf {\bibinfo
  {volume} {78}},\ \bibinfo {pages} {042307} (\bibinfo {year}
  {2008})}\BibitemShut {NoStop}%
\bibitem [{\citenamefont {Diehl}\ \emph {et~al.}(2008)\citenamefont {Diehl},
  \citenamefont {Micheli}, \citenamefont {Kantian}, \citenamefont {Kraus},
  \citenamefont {B\"uchler},\ and\ \citenamefont {Zoller}}]{Diehl08}%
  \BibitemOpen
  \bibfield  {author} {\bibinfo {author} {\bibfnamefont {S.}~\bibnamefont
  {Diehl}}, \bibinfo {author} {\bibfnamefont {A.}~\bibnamefont {Micheli}},
  \bibinfo {author} {\bibfnamefont {A.}~\bibnamefont {Kantian}}, \bibinfo
  {author} {\bibfnamefont {B.}~\bibnamefont {Kraus}}, \bibinfo {author}
  {\bibfnamefont {H.}~\bibnamefont {B\"uchler}}, \ and\ \bibinfo {author}
  {\bibfnamefont {P.}~\bibnamefont {Zoller}},\ }\href@noop {} {\bibfield
  {journal} {\bibinfo  {journal} {Nat. Phys}\ }\textbf {\bibinfo {volume}
  {4}},\ \bibinfo {pages} {878} (\bibinfo {year} {2008})}\BibitemShut {NoStop}%
\bibitem [{\citenamefont {Diehl}\ \emph {et~al.}(2011)\citenamefont {Diehl},
  \citenamefont {Rico}, \citenamefont {Baranov},\ and\ \citenamefont
  {Zoller}}]{Diehl11}%
  \BibitemOpen
  \bibfield  {author} {\bibinfo {author} {\bibfnamefont {S.}~\bibnamefont
  {Diehl}}, \bibinfo {author} {\bibfnamefont {E.}~\bibnamefont {Rico}},
  \bibinfo {author} {\bibfnamefont {M.~A.}\ \bibnamefont {Baranov}}, \ and\
  \bibinfo {author} {\bibfnamefont {P.}~\bibnamefont {Zoller}},\ }\href@noop {}
  {\bibfield  {journal} {\bibinfo  {journal} {Nat. Phys.}\ }\textbf {\bibinfo
  {volume} {7}},\ \bibinfo {pages} {971} (\bibinfo {year} {2011})}\BibitemShut
  {NoStop}%
\bibitem [{\citenamefont {Bardyn}\ \emph {et~al.}(2012)\citenamefont {Bardyn},
  \citenamefont {Baranov}, \citenamefont {Rico}, \citenamefont {\ifmmode
  \dot{I}\else \.{I}\fi{}mamo\ifmmode~\breve{g}\else \u{g}\fi{}lu},
  \citenamefont {Zoller},\ and\ \citenamefont {Diehl}}]{Bardyn12}%
  \BibitemOpen
  \bibfield  {author} {\bibinfo {author} {\bibfnamefont {C.-E.}\ \bibnamefont
  {Bardyn}}, \bibinfo {author} {\bibfnamefont {M.~A.}\ \bibnamefont {Baranov}},
  \bibinfo {author} {\bibfnamefont {E.}~\bibnamefont {Rico}}, \bibinfo {author}
  {\bibfnamefont {A.}~\bibnamefont {\ifmmode \dot{I}\else
  \.{I}\fi{}mamo\ifmmode~\breve{g}\else \u{g}\fi{}lu}}, \bibinfo {author}
  {\bibfnamefont {P.}~\bibnamefont {Zoller}}, \ and\ \bibinfo {author}
  {\bibfnamefont {S.}~\bibnamefont {Diehl}},\ }\href@noop {} {\bibfield
  {journal} {\bibinfo  {journal} {Phys. Rev. Lett.}\ }\textbf {\bibinfo
  {volume} {109}},\ \bibinfo {pages} {130402} (\bibinfo {year}
  {2012})}\BibitemShut {NoStop}%
\bibitem [{\citenamefont {Schindler}\ \emph {et~al.}(2013)\citenamefont
  {Schindler}, \citenamefont {M\"uller}, \citenamefont {Nigg}, \citenamefont
  {Barreiro}, \citenamefont {Mart\'inez}, \citenamefont {Hennrich},
  \citenamefont {Monz}, \citenamefont {Diehl}, \citenamefont {Zoller},\ and\
  \citenamefont {Blatt}}]{Schindler13}%
  \BibitemOpen
  \bibfield  {author} {\bibinfo {author} {\bibfnamefont {P.}~\bibnamefont
  {Schindler}}, \bibinfo {author} {\bibfnamefont {M.}~\bibnamefont {M\"uller}},
  \bibinfo {author} {\bibfnamefont {D.}~\bibnamefont {Nigg}}, \bibinfo {author}
  {\bibfnamefont {J.}~\bibnamefont {Barreiro}}, \bibinfo {author}
  {\bibfnamefont {E.}~\bibnamefont {Mart\'inez}}, \bibinfo {author}
  {\bibfnamefont {M.}~\bibnamefont {Hennrich}}, \bibinfo {author}
  {\bibfnamefont {T.}~\bibnamefont {Monz}}, \bibinfo {author} {\bibfnamefont
  {S.}~\bibnamefont {Diehl}}, \bibinfo {author} {\bibfnamefont
  {P.}~\bibnamefont {Zoller}}, \ and\ \bibinfo {author} {\bibfnamefont
  {R.}~\bibnamefont {Blatt}},\ }\href@noop {} {\bibfield  {journal} {\bibinfo
  {journal} {Nat. Phys}\ }\textbf {\bibinfo {volume} {9}},\ \bibinfo {pages}
  {361} (\bibinfo {year} {2013})}\BibitemShut {NoStop}%
\bibitem [{\citenamefont {Verstraete}\ \emph {et~al.}(2009)\citenamefont
  {Verstraete}, \citenamefont {Wolf},\ and\ \citenamefont
  {Cirac}}]{Verstraete09}%
  \BibitemOpen
  \bibfield  {author} {\bibinfo {author} {\bibfnamefont {F.}~\bibnamefont
  {Verstraete}}, \bibinfo {author} {\bibfnamefont {M.}~\bibnamefont {Wolf}}, \
  and\ \bibinfo {author} {\bibfnamefont {J.}~\bibnamefont {Cirac}},\
  }\href@noop {} {\bibfield  {journal} {\bibinfo  {journal} {Nat. Phys}\
  }\textbf {\bibinfo {volume} {5}},\ \bibinfo {pages} {633} (\bibinfo {year}
  {2009})}\BibitemShut {NoStop}%
\bibitem [{\citenamefont {Fredrickson}\ and\ \citenamefont
  {Andersen}(1984)}]{Fredrickson1984}%
  \BibitemOpen
  \bibfield  {author} {\bibinfo {author} {\bibfnamefont {G.~H.}\ \bibnamefont
  {Fredrickson}}\ and\ \bibinfo {author} {\bibfnamefont {H.~C.}\ \bibnamefont
  {Andersen}},\ }\href@noop {} {\bibfield  {journal} {\bibinfo  {journal}
  {Phys. Rev. Lett.}\ }\textbf {\bibinfo {volume} {53}},\ \bibinfo {pages}
  {1244} (\bibinfo {year} {1984})}\BibitemShut {NoStop}%
\bibitem [{\citenamefont {Jackle}\ and\ \citenamefont
  {Eisinger}(1991)}]{Jackle1991}%
  \BibitemOpen
  \bibfield  {author} {\bibinfo {author} {\bibfnamefont {J.}~\bibnamefont
  {Jackle}}\ and\ \bibinfo {author} {\bibfnamefont {S.}~\bibnamefont
  {Eisinger}},\ }\href@noop {} {\bibfield  {journal} {\bibinfo  {journal} {Z.
  Phys. B}\ }\textbf {\bibinfo {volume} {84}},\ \bibinfo {pages} {115}
  (\bibinfo {year} {1991})}\BibitemShut {NoStop}%
\bibitem [{\citenamefont {Ritort}\ and\ \citenamefont
  {Sollich}(2003)}]{Ritort2003}%
  \BibitemOpen
  \bibfield  {author} {\bibinfo {author} {\bibfnamefont {F.}~\bibnamefont
  {Ritort}}\ and\ \bibinfo {author} {\bibfnamefont {P.}~\bibnamefont
  {Sollich}},\ }\href@noop {} {\bibfield  {journal} {\bibinfo  {journal}
  {Advances in Physics}\ }\textbf {\bibinfo {volume} {52}},\ \bibinfo {pages}
  {219} (\bibinfo {year} {2003})}\BibitemShut {NoStop}%
\bibitem [{\citenamefont {Garrahan}\ \emph {et~al.}()\citenamefont {Garrahan},
  \citenamefont {Sollich},\ and\ \citenamefont {Toninelli}}]{Garrahan2010}%
  \BibitemOpen
  \bibfield  {author} {\bibinfo {author} {\bibfnamefont {J.~P.}\ \bibnamefont
  {Garrahan}}, \bibinfo {author} {\bibfnamefont {P.}~\bibnamefont {Sollich}}, \
  and\ \bibinfo {author} {\bibfnamefont {C.}~\bibnamefont {Toninelli}},\
  }\href@noop {} {\bibinfo  {journal} {arXiv:1009.6113}\ }\BibitemShut
  {NoStop}%
\bibitem [{\citenamefont {Markland}\ \emph {et~al.}(2011)\citenamefont
  {Markland}, \citenamefont {Morrone}, \citenamefont {Berne}, \citenamefont
  {Miyazaki}, \citenamefont {Rabani},\ and\ \citenamefont
  {Reichman}}]{Markland11}%
  \BibitemOpen
\bibfield  {journal} {  }\bibfield  {author} {\bibinfo {author} {\bibfnamefont
  {T.~E.}\ \bibnamefont {Markland}}, \bibinfo {author} {\bibfnamefont {J.~A.}\
  \bibnamefont {Morrone}}, \bibinfo {author} {\bibfnamefont {B.~J.}\
  \bibnamefont {Berne}}, \bibinfo {author} {\bibfnamefont {K.}~\bibnamefont
  {Miyazaki}}, \bibinfo {author} {\bibfnamefont {E.}~\bibnamefont {Rabani}}, \
  and\ \bibinfo {author} {\bibfnamefont {D.~R.}\ \bibnamefont {Reichman}},\
  }\href@noop {} {\bibfield  {journal} {\bibinfo  {journal} {Nat. Phys.}\
  }\textbf {\bibinfo {volume} {7}},\ \bibinfo {pages} {134} (\bibinfo {year}
  {2011})}\BibitemShut {NoStop}%
\bibitem [{\citenamefont {Olmos}\ \emph {et~al.}(2012)\citenamefont {Olmos},
  \citenamefont {Lesanovsky},\ and\ \citenamefont {Garrahan}}]{Olmos12}%
  \BibitemOpen
  \bibfield  {author} {\bibinfo {author} {\bibfnamefont {B.}~\bibnamefont
  {Olmos}}, \bibinfo {author} {\bibfnamefont {I.}~\bibnamefont {Lesanovsky}}, \
  and\ \bibinfo {author} {\bibfnamefont {J.~P.}\ \bibnamefont {Garrahan}},\
  }\href@noop {} {\bibfield  {journal} {\bibinfo  {journal} {Phys. Rev. Lett.}\
  }\textbf {\bibinfo {volume} {109}},\ \bibinfo {pages} {020403} (\bibinfo
  {year} {2012})}\BibitemShut {NoStop}%
\bibitem [{\citenamefont {Nussinov}\ \emph {et~al.}(2012)\citenamefont
  {Nussinov}, \citenamefont {Johnson}, \citenamefont {Graf},\ and\
  \citenamefont {Balatsky}}]{Nussinov2012}%
  \BibitemOpen
  \bibfield  {author} {\bibinfo {author} {\bibfnamefont {Z.}~\bibnamefont
  {Nussinov}}, \bibinfo {author} {\bibfnamefont {P.}~\bibnamefont {Johnson}},
  \bibinfo {author} {\bibfnamefont {M.~J.}\ \bibnamefont {Graf}}, \ and\
  \bibinfo {author} {\bibfnamefont {A.~V.}\ \bibnamefont {Balatsky}},\ }\href
  {http://arxiv.org/abs/1209.3823} {\  (\bibinfo {year} {2012})},\ \Eprint
  {http://arxiv.org/abs/1209.3823} {1209.3823} \BibitemShut {NoStop}%
\bibitem [{\citenamefont {Lechner}\ and\ \citenamefont
  {Zoller}(2013)}]{Lechner13}%
  \BibitemOpen
  \bibfield  {author} {\bibinfo {author} {\bibfnamefont {W.}~\bibnamefont
  {Lechner}}\ and\ \bibinfo {author} {\bibfnamefont {P.}~\bibnamefont
  {Zoller}},\ }\href@noop {} {\bibfield  {journal} {\bibinfo  {journal} {Phys.
  Rev. Lett.}\ }\textbf {\bibinfo {volume} {111}},\ \bibinfo {pages} {185306}
  (\bibinfo {year} {2013})}\BibitemShut {NoStop}%
\bibitem [{\citenamefont {{Diaz-Mendez}}\ \emph {et~al.}(2014)\citenamefont
  {{Diaz-Mendez}}, \citenamefont {{Mezzacapo}}, \citenamefont {{Cinti}},
  \citenamefont {{Lechner}},\ and\ \citenamefont
  {{Pupillo}}}]{Diaz-Mendez2014}%
  \BibitemOpen
  \bibfield  {author} {\bibinfo {author} {\bibfnamefont {R.}~\bibnamefont
  {{Diaz-Mendez}}}, \bibinfo {author} {\bibfnamefont {F.}~\bibnamefont
  {{Mezzacapo}}}, \bibinfo {author} {\bibfnamefont {F.}~\bibnamefont
  {{Cinti}}}, \bibinfo {author} {\bibfnamefont {W.}~\bibnamefont {{Lechner}}},
  \ and\ \bibinfo {author} {\bibfnamefont {G.}~\bibnamefont {{Pupillo}}},\
  }\href@noop {} {\  (\bibinfo {year} {2014})},\ \Eprint
  {http://arxiv.org/abs/1402.0852} {arXiv:1402.0852} \BibitemShut {NoStop}%
\bibitem [{\citenamefont {Rigol}\ \emph {et~al.}(2008)\citenamefont {Rigol},
  \citenamefont {Dunjko},\ and\ \citenamefont {Olshanii}}]{Rigol08}%
  \BibitemOpen
  \bibfield  {author} {\bibinfo {author} {\bibfnamefont {M.}~\bibnamefont
  {Rigol}}, \bibinfo {author} {\bibfnamefont {V.}~\bibnamefont {Dunjko}}, \
  and\ \bibinfo {author} {\bibfnamefont {M.}~\bibnamefont {Olshanii}},\
  }\href@noop {} {\bibfield  {journal} {\bibinfo  {journal} {Nature}\ }\textbf
  {\bibinfo {volume} {452}},\ \bibinfo {pages} {854} (\bibinfo {year}
  {2008})}\BibitemShut {NoStop}%
\bibitem [{\citenamefont {Polkovnikov}\ \emph {et~al.}(2011)\citenamefont
  {Polkovnikov}, \citenamefont {Sengupta}, \citenamefont {Silva},\ and\
  \citenamefont {Vengalattore}}]{Polkovnikov2011}%
  \BibitemOpen
  \bibfield  {author} {\bibinfo {author} {\bibfnamefont {A.}~\bibnamefont
  {Polkovnikov}}, \bibinfo {author} {\bibfnamefont {K.}~\bibnamefont
  {Sengupta}}, \bibinfo {author} {\bibfnamefont {A.}~\bibnamefont {Silva}}, \
  and\ \bibinfo {author} {\bibfnamefont {M.}~\bibnamefont {Vengalattore}},\
  }\href@noop {} {\bibfield  {journal} {\bibinfo  {journal} {Rev. Mod. Phys.}\
  }\textbf {\bibinfo {volume} {83}},\ \bibinfo {pages} {863} (\bibinfo {year}
  {2011})}\BibitemShut {NoStop}%
\bibitem [{\citenamefont {Basko}\ \emph {et~al.}(2006)\citenamefont {Basko},
  \citenamefont {Aleiner},\ and\ \citenamefont {Altshuler}}]{Basko2006}%
  \BibitemOpen
  \bibfield  {author} {\bibinfo {author} {\bibfnamefont {D.}~\bibnamefont
  {Basko}}, \bibinfo {author} {\bibfnamefont {I.}~\bibnamefont {Aleiner}}, \
  and\ \bibinfo {author} {\bibfnamefont {B.}~\bibnamefont {Altshuler}},\
  }\href@noop {} {\bibfield  {journal} {\bibinfo  {journal} {Ann. Phys.}\
  }\textbf {\bibinfo {volume} {321}},\ \bibinfo {pages} {1126} (\bibinfo {year}
  {2006})}\BibitemShut {NoStop}%
\bibitem [{\citenamefont {Pal}\ and\ \citenamefont {Huse}(2010)}]{Pal2010}%
  \BibitemOpen
  \bibfield  {author} {\bibinfo {author} {\bibfnamefont {A.}~\bibnamefont
  {Pal}}\ and\ \bibinfo {author} {\bibfnamefont {D.~A.}\ \bibnamefont {Huse}},\
  }\href@noop {} {\bibfield  {journal} {\bibinfo  {journal} {Phys. Rev. B}\
  }\textbf {\bibinfo {volume} {82}},\ \bibinfo {pages} {174411} (\bibinfo
  {year} {2010})}\BibitemShut {NoStop}%
\bibitem [{\citenamefont {{Nandkishore}}\ and\ \citenamefont
  {{Huse}}(2014)}]{Nandkishore2014}%
  \BibitemOpen
  \bibfield  {author} {\bibinfo {author} {\bibfnamefont {R.}~\bibnamefont
  {{Nandkishore}}}\ and\ \bibinfo {author} {\bibfnamefont {D.~A.}\ \bibnamefont
  {{Huse}}},\ }\href@noop {} {\  (\bibinfo {year} {2014})},\ \Eprint
  {http://arxiv.org/abs/1404.0686} {arXiv:1404.0686} \BibitemShut {NoStop}%
\bibitem [{\citenamefont {Gallagher}(1984)}]{Gallagher84}%
  \BibitemOpen
  \bibfield  {author} {\bibinfo {author} {\bibfnamefont {T.}~\bibnamefont
  {Gallagher}},\ }\href@noop {} {\emph {\bibinfo {title} {Rydberg Atoms}}}\
  (\bibinfo  {publisher} {Cambridge University Press},\ \bibinfo {year}
  {1984})\BibitemShut {NoStop}%
\bibitem [{\citenamefont {Pritchard}\ \emph {et~al.}(2010)\citenamefont
  {Pritchard}, \citenamefont {Maxwell}, \citenamefont {Gauguet}, \citenamefont
  {Weatherill}, \citenamefont {Jones},\ and\ \citenamefont
  {Adams}}]{Pritchard10}%
  \BibitemOpen
  \bibfield  {author} {\bibinfo {author} {\bibfnamefont {J.~D.}\ \bibnamefont
  {Pritchard}}, \bibinfo {author} {\bibfnamefont {D.}~\bibnamefont {Maxwell}},
  \bibinfo {author} {\bibfnamefont {A.}~\bibnamefont {Gauguet}}, \bibinfo
  {author} {\bibfnamefont {K.~J.}\ \bibnamefont {Weatherill}}, \bibinfo
  {author} {\bibfnamefont {M.~P.~A.}\ \bibnamefont {Jones}}, \ and\ \bibinfo
  {author} {\bibfnamefont {C.~S.}\ \bibnamefont {Adams}},\ }\href {\doibase
  10.1103/PhysRevLett.105.193603} {\bibfield  {journal} {\bibinfo  {journal}
  {Phys. Rev. Lett.}\ }\textbf {\bibinfo {volume} {105}},\ \bibinfo {pages}
  {193603} (\bibinfo {year} {2010})}\BibitemShut {NoStop}%
\bibitem [{\citenamefont {Adams}\ \emph {et~al.}(2013)\citenamefont {Adams},
  \citenamefont {Weatherill},\ and\ \citenamefont {Pritchard}}]{Adams13}%
  \BibitemOpen
  \bibfield  {author} {\bibinfo {author} {\bibfnamefont {C.~S.}\ \bibnamefont
  {Adams}}, \bibinfo {author} {\bibfnamefont {K.~J.}\ \bibnamefont
  {Weatherill}}, \ and\ \bibinfo {author} {\bibfnamefont {J.~D.}\ \bibnamefont
  {Pritchard}},\ }\enquote {\bibinfo {title} {Nonlinear optics using cold
  rydberg atoms},}\ in\ \href@noop {} {\emph {\bibinfo {booktitle} {Annual
  Review of Cold Atoms and Molecules}}}\ (\bibinfo {year} {2013})\
  Chap.~\bibinfo {chapter} {8}, pp.\ \bibinfo {pages} {301--350}\BibitemShut
  {NoStop}%
\bibitem [{\citenamefont {Schempp}\ \emph {et~al.}(2010)\citenamefont
  {Schempp}, \citenamefont {G\"unter}, \citenamefont {Hofmann}, \citenamefont
  {Giese}, \citenamefont {Saliba}, \citenamefont {DePaola}, \citenamefont
  {Amthor}, \citenamefont {Weidem\"uller}, \citenamefont
  {Sevin\ifmmode~\mbox{\c{c}}\else \c{c}\fi{}li},\ and\ \citenamefont
  {Pohl}}]{Schempp10}%
  \BibitemOpen
  \bibfield  {author} {\bibinfo {author} {\bibfnamefont {H.}~\bibnamefont
  {Schempp}}, \bibinfo {author} {\bibfnamefont {G.}~\bibnamefont {G\"unter}},
  \bibinfo {author} {\bibfnamefont {C.~S.}\ \bibnamefont {Hofmann}}, \bibinfo
  {author} {\bibfnamefont {C.}~\bibnamefont {Giese}}, \bibinfo {author}
  {\bibfnamefont {S.~D.}\ \bibnamefont {Saliba}}, \bibinfo {author}
  {\bibfnamefont {B.~D.}\ \bibnamefont {DePaola}}, \bibinfo {author}
  {\bibfnamefont {T.}~\bibnamefont {Amthor}}, \bibinfo {author} {\bibfnamefont
  {M.}~\bibnamefont {Weidem\"uller}}, \bibinfo {author} {\bibfnamefont
  {S.}~\bibnamefont {Sevin\ifmmode~\mbox{\c{c}}\else \c{c}\fi{}li}}, \ and\
  \bibinfo {author} {\bibfnamefont {T.}~\bibnamefont {Pohl}},\ }\href@noop {}
  {\bibfield  {journal} {\bibinfo  {journal} {Phys. Rev. Lett.}\ }\textbf
  {\bibinfo {volume} {104}},\ \bibinfo {pages} {173602} (\bibinfo {year}
  {2010})}\BibitemShut {NoStop}%
\bibitem [{\citenamefont {Schwarzkopf}\ \emph {et~al.}(2011)\citenamefont
  {Schwarzkopf}, \citenamefont {Sapiro},\ and\ \citenamefont
  {Raithel}}]{Schwarzkopf11}%
  \BibitemOpen
  \bibfield  {author} {\bibinfo {author} {\bibfnamefont {A.}~\bibnamefont
  {Schwarzkopf}}, \bibinfo {author} {\bibfnamefont {R.~E.}\ \bibnamefont
  {Sapiro}}, \ and\ \bibinfo {author} {\bibfnamefont {G.}~\bibnamefont
  {Raithel}},\ }\href {\doibase 10.1103/PhysRevLett.107.103001} {\bibfield
  {journal} {\bibinfo  {journal} {Phys. Rev. Lett.}\ }\textbf {\bibinfo
  {volume} {107}},\ \bibinfo {pages} {103001} (\bibinfo {year}
  {2011})}\BibitemShut {NoStop}%
\bibitem [{\citenamefont {Schwarzkopf}\ \emph {et~al.}(2013)\citenamefont
  {Schwarzkopf}, \citenamefont {Anderson}, \citenamefont {Thaicharoen},\ and\
  \citenamefont {Raithel}}]{Schwarzkopf13}%
  \BibitemOpen
  \bibfield  {author} {\bibinfo {author} {\bibfnamefont {A.}~\bibnamefont
  {Schwarzkopf}}, \bibinfo {author} {\bibfnamefont {D.~A.}\ \bibnamefont
  {Anderson}}, \bibinfo {author} {\bibfnamefont {N.}~\bibnamefont
  {Thaicharoen}}, \ and\ \bibinfo {author} {\bibfnamefont {G.}~\bibnamefont
  {Raithel}},\ }\href {\doibase 10.1103/PhysRevA.88.061406} {\bibfield
  {journal} {\bibinfo  {journal} {Phys. Rev. A}\ }\textbf {\bibinfo {volume}
  {88}},\ \bibinfo {pages} {061406} (\bibinfo {year} {2013})}\BibitemShut
  {NoStop}%
\bibitem [{\citenamefont {Peyronel}\ \emph {et~al.}(2012)\citenamefont
  {Peyronel}, \citenamefont {Firstenberg}, \citenamefont {Liang}, \citenamefont
  {Hofferberth}, \citenamefont {Gorshkov}, \citenamefont {Pohl}, \citenamefont
  {Lukin},\ and\ \citenamefont {Vuletic}}]{Peyronel12}%
  \BibitemOpen
  \bibfield  {author} {\bibinfo {author} {\bibfnamefont {T.}~\bibnamefont
  {Peyronel}}, \bibinfo {author} {\bibfnamefont {O.}~\bibnamefont
  {Firstenberg}}, \bibinfo {author} {\bibfnamefont {Q.-Y.}\ \bibnamefont
  {Liang}}, \bibinfo {author} {\bibfnamefont {S.}~\bibnamefont {Hofferberth}},
  \bibinfo {author} {\bibfnamefont {A.~V.}\ \bibnamefont {Gorshkov}}, \bibinfo
  {author} {\bibfnamefont {T.}~\bibnamefont {Pohl}}, \bibinfo {author}
  {\bibfnamefont {M.~D.}\ \bibnamefont {Lukin}}, \ and\ \bibinfo {author}
  {\bibfnamefont {V.}~\bibnamefont {Vuletic}},\ }\href@noop {} {\bibfield
  {journal} {\bibinfo  {journal} {Nature}\ }\textbf {\bibinfo {volume} {488}},\
  \bibinfo {pages} {57} (\bibinfo {year} {2012})}\BibitemShut {NoStop}%
\bibitem [{\citenamefont {Maxwell}\ \emph {et~al.}(2013)\citenamefont
  {Maxwell}, \citenamefont {Szwer}, \citenamefont {Paredes-Barato},
  \citenamefont {Busche}, \citenamefont {Pritchard}, \citenamefont {Gauguet},
  \citenamefont {Weatherill}, \citenamefont {Jones},\ and\ \citenamefont
  {Adams}}]{Maxwell13}%
  \BibitemOpen
  \bibfield  {author} {\bibinfo {author} {\bibfnamefont {D.}~\bibnamefont
  {Maxwell}}, \bibinfo {author} {\bibfnamefont {D.~J.}\ \bibnamefont {Szwer}},
  \bibinfo {author} {\bibfnamefont {D.}~\bibnamefont {Paredes-Barato}},
  \bibinfo {author} {\bibfnamefont {H.}~\bibnamefont {Busche}}, \bibinfo
  {author} {\bibfnamefont {J.~D.}\ \bibnamefont {Pritchard}}, \bibinfo {author}
  {\bibfnamefont {A.}~\bibnamefont {Gauguet}}, \bibinfo {author} {\bibfnamefont
  {K.~J.}\ \bibnamefont {Weatherill}}, \bibinfo {author} {\bibfnamefont
  {M.~P.~A.}\ \bibnamefont {Jones}}, \ and\ \bibinfo {author} {\bibfnamefont
  {C.~S.}\ \bibnamefont {Adams}},\ }\href@noop {} {\bibfield  {journal}
  {\bibinfo  {journal} {Phys. Rev. Lett.}\ }\textbf {\bibinfo {volume} {110}},\
  \bibinfo {pages} {103001} (\bibinfo {year} {2013})}\BibitemShut {NoStop}%
\bibitem [{\citenamefont {Ates}\ \emph {et~al.}(2006)\citenamefont {Ates},
  \citenamefont {Pohl}, \citenamefont {Pattard},\ and\ \citenamefont
  {Rost}}]{Ates06}%
  \BibitemOpen
  \bibfield  {author} {\bibinfo {author} {\bibfnamefont {C.}~\bibnamefont
  {Ates}}, \bibinfo {author} {\bibfnamefont {T.}~\bibnamefont {Pohl}}, \bibinfo
  {author} {\bibfnamefont {T.}~\bibnamefont {Pattard}}, \ and\ \bibinfo
  {author} {\bibfnamefont {J.~M.}\ \bibnamefont {Rost}},\ }\href@noop {}
  {\bibfield  {journal} {\bibinfo  {journal} {Journal of Physics B}\ }\textbf
  {\bibinfo {volume} {39}},\ \bibinfo {pages} {L233} (\bibinfo {year}
  {2006})}\BibitemShut {NoStop}%
\bibitem [{\citenamefont {Ates}\ \emph
  {et~al.}(2007{\natexlab{a}})\citenamefont {Ates}, \citenamefont {Pohl},
  \citenamefont {Pattard},\ and\ \citenamefont {Rost}}]{Ates07-1}%
  \BibitemOpen
  \bibfield  {author} {\bibinfo {author} {\bibfnamefont {C.}~\bibnamefont
  {Ates}}, \bibinfo {author} {\bibfnamefont {T.}~\bibnamefont {Pohl}}, \bibinfo
  {author} {\bibfnamefont {T.}~\bibnamefont {Pattard}}, \ and\ \bibinfo
  {author} {\bibfnamefont {J.~M.}\ \bibnamefont {Rost}},\ }\href@noop {}
  {\bibfield  {journal} {\bibinfo  {journal} {Phys. Rev. Lett.}\ }\textbf
  {\bibinfo {volume} {98}},\ \bibinfo {pages} {023002} (\bibinfo {year}
  {2007}{\natexlab{a}})}\BibitemShut {NoStop}%
\bibitem [{\citenamefont {Ates}\ \emph
  {et~al.}(2007{\natexlab{b}})\citenamefont {Ates}, \citenamefont {Pohl},
  \citenamefont {Pattard},\ and\ \citenamefont {Rost}}]{Ates07-2}%
  \BibitemOpen
  \bibfield  {author} {\bibinfo {author} {\bibfnamefont {C.}~\bibnamefont
  {Ates}}, \bibinfo {author} {\bibfnamefont {T.}~\bibnamefont {Pohl}}, \bibinfo
  {author} {\bibfnamefont {T.}~\bibnamefont {Pattard}}, \ and\ \bibinfo
  {author} {\bibfnamefont {J.~M.}\ \bibnamefont {Rost}},\ }\href@noop {}
  {\bibfield  {journal} {\bibinfo  {journal} {Phys. Rev. A}\ }\textbf {\bibinfo
  {volume} {76}},\ \bibinfo {pages} {013413} (\bibinfo {year}
  {2007}{\natexlab{b}})}\BibitemShut {NoStop}%
\bibitem [{\citenamefont {H\"oning}\ \emph {et~al.}(2013)\citenamefont
  {H\"oning}, \citenamefont {Muth}, \citenamefont {Petrosyan},\ and\
  \citenamefont {Fleischhauer}}]{Hoening13}%
  \BibitemOpen
  \bibfield  {author} {\bibinfo {author} {\bibfnamefont {M.}~\bibnamefont
  {H\"oning}}, \bibinfo {author} {\bibfnamefont {D.}~\bibnamefont {Muth}},
  \bibinfo {author} {\bibfnamefont {D.}~\bibnamefont {Petrosyan}}, \ and\
  \bibinfo {author} {\bibfnamefont {M.}~\bibnamefont {Fleischhauer}},\
  }\href@noop {} {\bibfield  {journal} {\bibinfo  {journal} {Phys. Rev. A}\
  }\textbf {\bibinfo {volume} {87}},\ \bibinfo {pages} {023401} (\bibinfo
  {year} {2013})}\BibitemShut {NoStop}%
\bibitem [{\citenamefont {Petrosyan}\ \emph {et~al.}(2013)\citenamefont
  {Petrosyan}, \citenamefont {H\"oning},\ and\ \citenamefont
  {Fleischhauer}}]{Petrosyan13}%
  \BibitemOpen
  \bibfield  {author} {\bibinfo {author} {\bibfnamefont {D.}~\bibnamefont
  {Petrosyan}}, \bibinfo {author} {\bibfnamefont {M.}~\bibnamefont {H\"oning}},
  \ and\ \bibinfo {author} {\bibfnamefont {M.}~\bibnamefont {Fleischhauer}},\
  }\href@noop {} {\bibfield  {journal} {\bibinfo  {journal} {Phys. Rev. A}\
  }\textbf {\bibinfo {volume} {87}},\ \bibinfo {pages} {053414} (\bibinfo
  {year} {2013})}\BibitemShut {NoStop}%
\bibitem [{\citenamefont {Petrosyan}(2013)}]{Petrosyan13-1}%
  \BibitemOpen
  \bibfield  {author} {\bibinfo {author} {\bibfnamefont {D.}~\bibnamefont
  {Petrosyan}},\ }\href@noop {} {\bibfield  {journal} {\bibinfo  {journal}
  {Phys. Rev. A}\ }\textbf {\bibinfo {volume} {88}},\ \bibinfo {pages} {043431}
  (\bibinfo {year} {2013})}\BibitemShut {NoStop}%
\bibitem [{\citenamefont {Hoening}\ \emph {et~al.}(2014)\citenamefont
  {Hoening}, \citenamefont {Abdussalam}, \citenamefont {Fleischhauer},\ and\
  \citenamefont {Pohl}}]{Hoening14}%
  \BibitemOpen
  \bibfield  {author} {\bibinfo {author} {\bibfnamefont {M.}~\bibnamefont
  {Hoening}}, \bibinfo {author} {\bibfnamefont {W.}~\bibnamefont {Abdussalam}},
  \bibinfo {author} {\bibfnamefont {M.}~\bibnamefont {Fleischhauer}}, \ and\
  \bibinfo {author} {\bibfnamefont {T.}~\bibnamefont {Pohl}},\ }\href@noop {}
  {\bibfield  {journal} {\bibinfo  {journal} {preprint}\ ,\ \bibinfo {pages}
  {arXiv:1404.1281}} (\bibinfo {year} {2014})}\BibitemShut {NoStop}%
\bibitem [{\citenamefont {Sanders}\ \emph {et~al.}(2014)\citenamefont
  {Sanders}, \citenamefont {van Bijnen}, \citenamefont {Vredenbregt},\ and\
  \citenamefont {Kokkelmans}}]{Sanders14}%
  \BibitemOpen
  \bibfield  {author} {\bibinfo {author} {\bibfnamefont {J.}~\bibnamefont
  {Sanders}}, \bibinfo {author} {\bibfnamefont {R.}~\bibnamefont {van Bijnen}},
  \bibinfo {author} {\bibfnamefont {E.}~\bibnamefont {Vredenbregt}}, \ and\
  \bibinfo {author} {\bibfnamefont {S.}~\bibnamefont {Kokkelmans}},\
  }\href@noop {} {\bibfield  {journal} {\bibinfo  {journal} {Phys. Rev. Lett.}\
  }\textbf {\bibinfo {volume} {112}},\ \bibinfo {pages} {163001} (\bibinfo
  {year} {2014})}\BibitemShut {NoStop}%
\bibitem [{\citenamefont {Gardiner}\ and\ \citenamefont
  {Zoller}(2004)}]{Gardiner04}%
  \BibitemOpen
  \bibfield  {author} {\bibinfo {author} {\bibfnamefont {C.~W.}\ \bibnamefont
  {Gardiner}}\ and\ \bibinfo {author} {\bibfnamefont {P.}~\bibnamefont
  {Zoller}},\ }\href@noop {} {\emph {\bibinfo {title} {Quantum Noise}}}\
  (\bibinfo  {publisher} {Springer-Verlag},\ \bibinfo {year}
  {2004})\BibitemShut {NoStop}%
\bibitem [{\citenamefont {M\o{}lmer}\ \emph {et~al.}(1993)\citenamefont
  {M\o{}lmer}, \citenamefont {Castin},\ and\ \citenamefont
  {Dalibard}}]{Molmer93}%
  \BibitemOpen
  \bibfield  {author} {\bibinfo {author} {\bibfnamefont {K.}~\bibnamefont
  {M\o{}lmer}}, \bibinfo {author} {\bibfnamefont {Y.}~\bibnamefont {Castin}}, \
  and\ \bibinfo {author} {\bibfnamefont {J.}~\bibnamefont {Dalibard}},\
  }\href@noop {} {\bibfield  {journal} {\bibinfo  {journal} {J. Opt. Soc. Am.
  B}\ }\textbf {\bibinfo {volume} {10}},\ \bibinfo {pages} {524} (\bibinfo
  {year} {1993})}\BibitemShut {NoStop}%
\bibitem [{\citenamefont {Dalibard}\ \emph {et~al.}(1992)\citenamefont
  {Dalibard}, \citenamefont {Castin},\ and\ \citenamefont
  {M\o{}lmer}}]{Dalibard92}%
  \BibitemOpen
  \bibfield  {author} {\bibinfo {author} {\bibfnamefont {J.}~\bibnamefont
  {Dalibard}}, \bibinfo {author} {\bibfnamefont {Y.}~\bibnamefont {Castin}}, \
  and\ \bibinfo {author} {\bibfnamefont {K.}~\bibnamefont {M\o{}lmer}},\
  }\href@noop {} {\bibfield  {journal} {\bibinfo  {journal} {Phys. Rev. Lett.}\
  }\textbf {\bibinfo {volume} {68}},\ \bibinfo {pages} {580} (\bibinfo {year}
  {1992})}\BibitemShut {NoStop}%
\bibitem [{Note1()}]{Note1}%
  \BibitemOpen
  \bibinfo {note} {Note, that there are other ways to define quantum versions
  of these KCMs which are not purely dissipative \cite {Olmos12}.}\BibitemShut
  {Stop}%
\bibitem [{\citenamefont {Chleboun}\ \emph {et~al.}(2013)\citenamefont
  {Chleboun}, \citenamefont {Faggionato},\ and\ \citenamefont
  {Martinelli}}]{Chleboun2013}%
  \BibitemOpen
  \bibfield  {author} {\bibinfo {author} {\bibfnamefont {P.}~\bibnamefont
  {Chleboun}}, \bibinfo {author} {\bibfnamefont {A.}~\bibnamefont
  {Faggionato}}, \ and\ \bibinfo {author} {\bibfnamefont {F.}~\bibnamefont
  {Martinelli}},\ }\href@noop {} {\bibfield  {journal} {\bibinfo  {journal} {J.
  Stat. Mech.}\ ,\ \bibinfo {pages} {L04001}} (\bibinfo {year}
  {2013})}\BibitemShut {NoStop}%
\bibitem [{\citenamefont {Sollich}\ and\ \citenamefont
  {Evans}(1999)}]{Sollich99}%
  \BibitemOpen
  \bibfield  {author} {\bibinfo {author} {\bibfnamefont {P.}~\bibnamefont
  {Sollich}}\ and\ \bibinfo {author} {\bibfnamefont {M.~R.}\ \bibnamefont
  {Evans}},\ }\href@noop {} {\bibfield  {journal} {\bibinfo  {journal} {Phys.
  Rev. Lett.}\ }\textbf {\bibinfo {volume} {83}},\ \bibinfo {pages} {3238}
  (\bibinfo {year} {1999})}\BibitemShut {NoStop}%
\bibitem [{\citenamefont {Lesanovsky}(2011)}]{Lesanovsky11}%
  \BibitemOpen
  \bibfield  {author} {\bibinfo {author} {\bibfnamefont {I.}~\bibnamefont
  {Lesanovsky}},\ }\href {\doibase 10.1103/PhysRevLett.106.025301} {\bibfield
  {journal} {\bibinfo  {journal} {Phys. Rev. Lett.}\ }\textbf {\bibinfo
  {volume} {106}},\ \bibinfo {pages} {025301} (\bibinfo {year}
  {2011})}\BibitemShut {NoStop}%
\bibitem [{\citenamefont {Lesanovsky}\ and\ \citenamefont
  {Garrahan}(2013)}]{Lesanovsky13}%
  \BibitemOpen
  \bibfield  {author} {\bibinfo {author} {\bibfnamefont {I.}~\bibnamefont
  {Lesanovsky}}\ and\ \bibinfo {author} {\bibfnamefont {J.~P.}\ \bibnamefont
  {Garrahan}},\ }\href@noop {} {\bibfield  {journal} {\bibinfo  {journal}
  {Phys. Rev. Lett.}\ }\textbf {\bibinfo {volume} {111}},\ \bibinfo {pages}
  {215305} (\bibinfo {year} {2013})}\BibitemShut {NoStop}%
\bibitem [{\citenamefont {Marcuzzi}\ \emph {et~al.}(2014)\citenamefont
  {Marcuzzi}, \citenamefont {Schick}, \citenamefont {Olmos},\ and\
  \citenamefont {Lesanovsky}}]{Marcuzzi14}%
  \BibitemOpen
  \bibfield  {author} {\bibinfo {author} {\bibfnamefont {M.}~\bibnamefont
  {Marcuzzi}}, \bibinfo {author} {\bibfnamefont {J.}~\bibnamefont {Schick}},
  \bibinfo {author} {\bibfnamefont {B.}~\bibnamefont {Olmos}}, \ and\ \bibinfo
  {author} {\bibfnamefont {I.}~\bibnamefont {Lesanovsky}},\ }\href@noop {} {\
  ,\ \bibinfo {pages} {In preparation} (\bibinfo {year} {2014})}\BibitemShut
  {NoStop}%
\bibitem [{\citenamefont {Diehl}\ \emph {et~al.}(2010)\citenamefont {Diehl},
  \citenamefont {Yi}, \citenamefont {Daley},\ and\ \citenamefont
  {Zoller}}]{Diehl10}%
  \BibitemOpen
  \bibfield  {author} {\bibinfo {author} {\bibfnamefont {S.}~\bibnamefont
  {Diehl}}, \bibinfo {author} {\bibfnamefont {W.}~\bibnamefont {Yi}}, \bibinfo
  {author} {\bibfnamefont {A.~J.}\ \bibnamefont {Daley}}, \ and\ \bibinfo
  {author} {\bibfnamefont {P.}~\bibnamefont {Zoller}},\ }\href@noop {}
  {\bibfield  {journal} {\bibinfo  {journal} {Phys. Rev. Lett.}\ }\textbf
  {\bibinfo {volume} {105}},\ \bibinfo {pages} {227001} (\bibinfo {year}
  {2010})}\BibitemShut {NoStop}%
\bibitem [{\citenamefont {Yi}\ \emph {et~al.}(2012)\citenamefont {Yi},
  \citenamefont {Diehl}, \citenamefont {Daley},\ and\ \citenamefont
  {Zoller}}]{Yi12}%
  \BibitemOpen
  \bibfield  {author} {\bibinfo {author} {\bibfnamefont {W.}~\bibnamefont
  {Yi}}, \bibinfo {author} {\bibfnamefont {S.}~\bibnamefont {Diehl}}, \bibinfo
  {author} {\bibfnamefont {A.~J.}\ \bibnamefont {Daley}}, \ and\ \bibinfo
  {author} {\bibfnamefont {P.}~\bibnamefont {Zoller}},\ }\href@noop {}
  {\bibfield  {journal} {\bibinfo  {journal} {New Journal of Physics}\ }\textbf
  {\bibinfo {volume} {14}},\ \bibinfo {pages} {055002} (\bibinfo {year}
  {2012})}\BibitemShut {NoStop}%
\end{thebibliography}
\end{document}